\documentclass{PoS}

\title{Kinematical correlations: from RHIC to LHC}

\ShortTitle{Kinematical correlations}

\author{\speaker{Antoni Szczurek}\\
        Institute of Nuclear Physics, PL-31-342 Cracow, Poland\\
        and Univeristy of Rzesz\'ow, PL-35-959 Cracow, Poland\\
        E-mail: \email{Antoni.Szczurek@ifj.edu.pl}}

\abstract{Kinematical correlations between outgoing 
partons (elementary particles) are discussed in 
the framework of $k_t$-factorization.
Different unintegrated gluon distributions are used.
I discuss correlations between charm-anticharm, jets, 
photon and jet, and between leptons from semileptonic
decays of heavy quarks as well as leptons produced
in the Drell-Yan mechanism. Both correlations in azimuth
as well as in the length of transverse momenta are shown.
General conclusions are drawn.}

\FullConference{High-pT Physics at LHC - Tokaj'08 \\
		 16-19 March 2008\\
		 Tokaj, Hungary}

\begin{document} 

\section{Introduction}

Most of the QCD calculations in the literature
concentrated on calculating inclusive distributions 
(heavy quarks, jets, direct photons, etc). The standard
collinear approach with next-to-leading order (NLO) 
accuracy is state of art in this respect.
Higher luminosity and in the consequence better statistics
at present and future colliders
give a new possibility to study not only inclusive
distributions but also correlations between outgoing
partons or other elementary particles 
(photons, leptons, etc.). The standard collinear
calculations are not the best suited for this purpose.
In particular, one encounters singularities for
back-to-back kinematics, i.e. when two elementary
particles are emitted at angles differing in azimuth
by 180$^o$ or having the same length of the transverse
momenta. At such configurations the standard collinear
approach is not reliable and other methods should be used.
The singularities can be avoided when initial transverse
momenta of partons are included. In this presentation
I discuss some general aspects within $k_t$-factorization
approach, where so-called unintegrated
gluon or parton distributions are used instead of
usual standard distributions. The UGDFs (or UPDFs) are
not only functions of longitudinal momentum fraction
but also of (initial) transverse momenta.
In general, there are two sources of initial transverse
momenta. One is internal (Fermi) motion of hadron
constituents. This, highly nonperturbative, effect
is very difficult to calculate from first principles
and therefore often neglected.
The second is a perturbative effect related to parton
evolution in QCD including $k_t$-kicks. The standard collinear 
approach limits itself somewhat arbitrarily to longitudinal 
directions of initial partons and neglects transverse degree 
of freedom. 

Our group has calculated correlation observables for several processes:
\begin{itemize}
\item $c \bar c$ correlations \cite{LS06}, 
\item dijet correlations \cite{SRS07}, 
\item photon-jet correlations \cite{PS07}
\item correlations of Drell-Yan dileptons and correlations of dilepton-jet \cite{SS08}.
\end{itemize}
In this talk I review the general strategy. The details can be found 
in original papers \cite{LS06,SRS07,PS07,SS08}. 
Only some selected detailed results are shown here. 
Some other results for $c \bar c$ and $e^+ e^-$ correlations were 
presented in parallel talks of my collaborators
(see \cite{Luszczak_thisproceedings,Slipek_thisproceedings}) 
at this workshop.

\section{Sketch of the formalism}


\begin{figure}    
\begin{center}
\includegraphics[width=5.0cm]{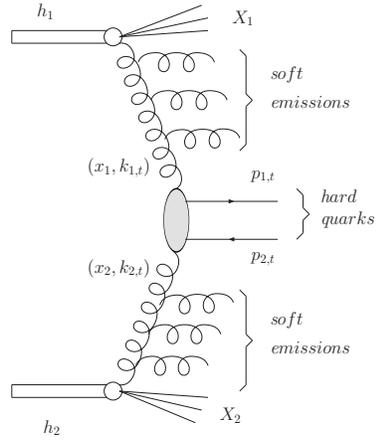}
\caption{Diagram for $k_t$-factorization approach to 
$c \bar c$ production.}
\label{fig:kt_factorization_ccbar_diagrams}
\end{center}
\end{figure}


\begin{figure}    
\begin{center}
\resizebox{0.35\columnwidth}{!}{%
\includegraphics{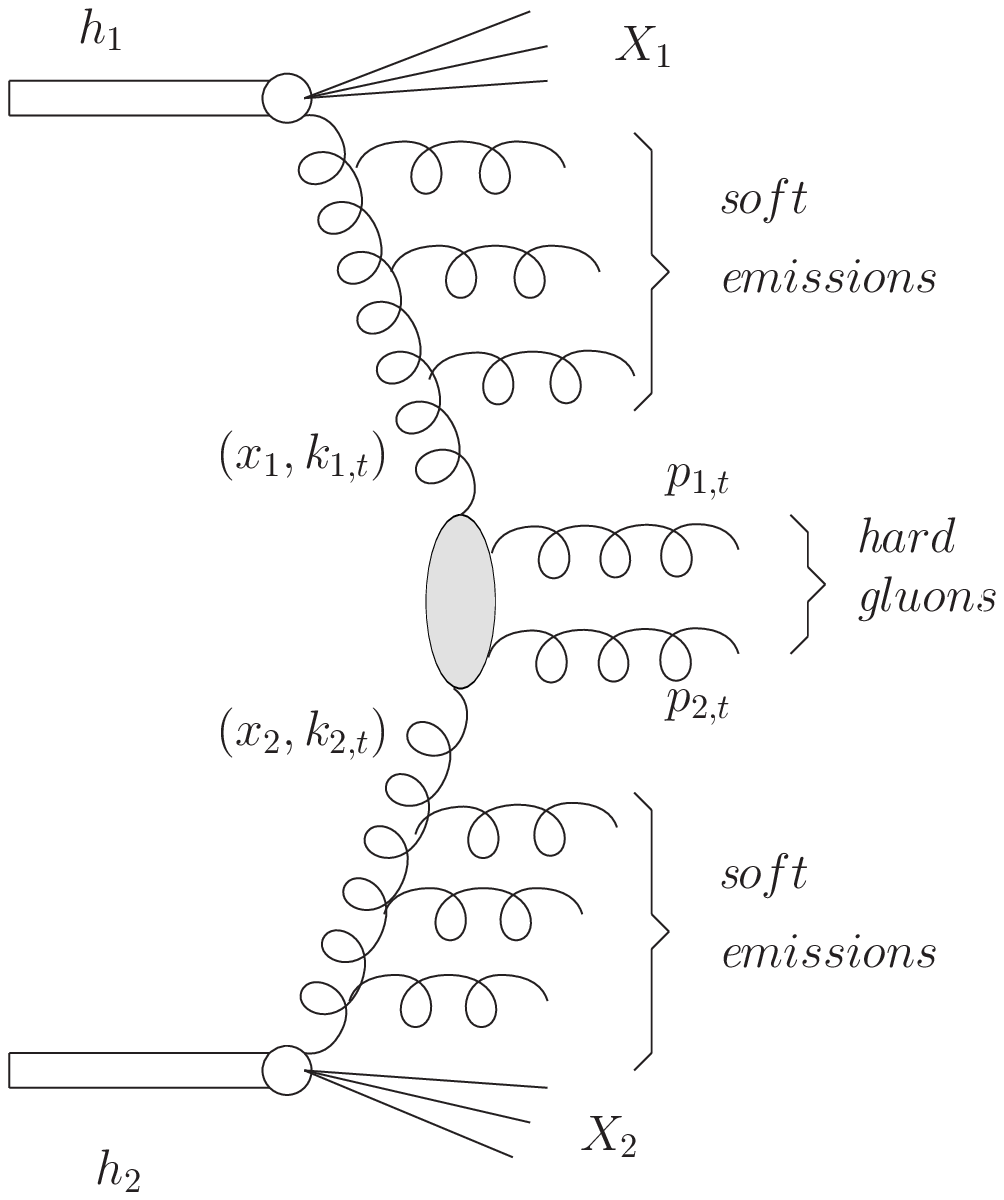} }
\resizebox{0.35\columnwidth}{!}{%
\includegraphics{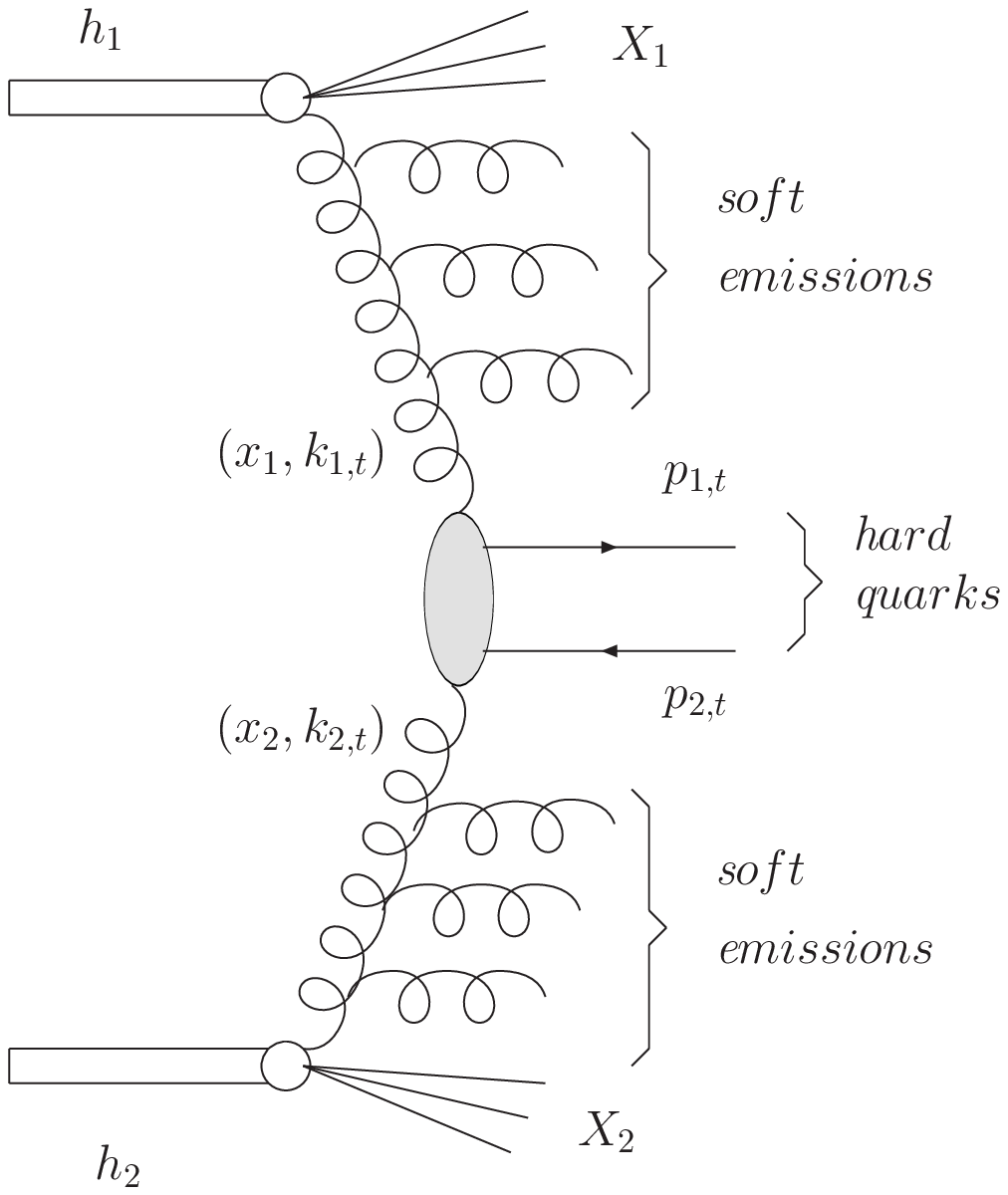} }
\resizebox{0.35\columnwidth}{!}{%
\includegraphics{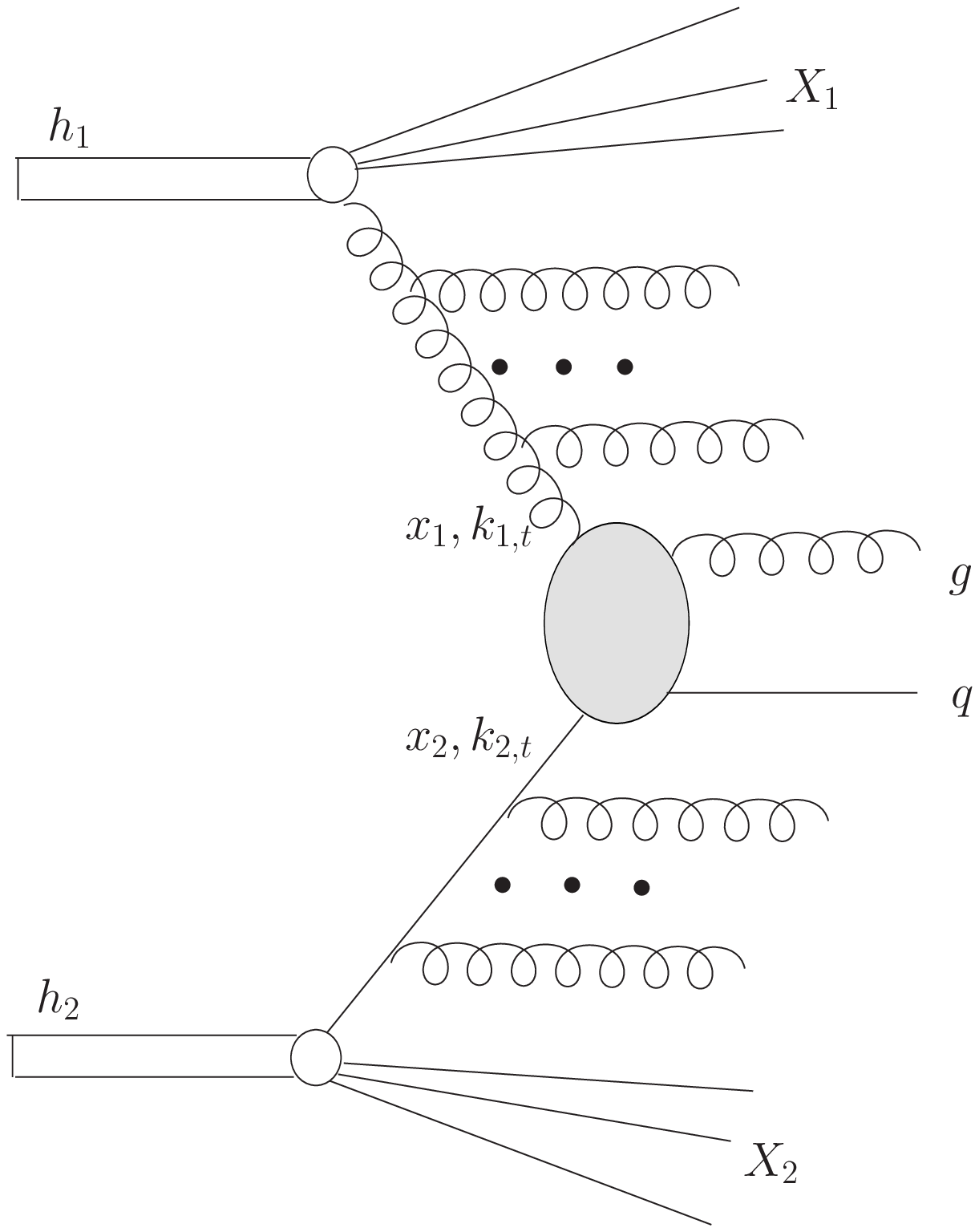} }
\resizebox{0.35\columnwidth}{!}{%
\includegraphics{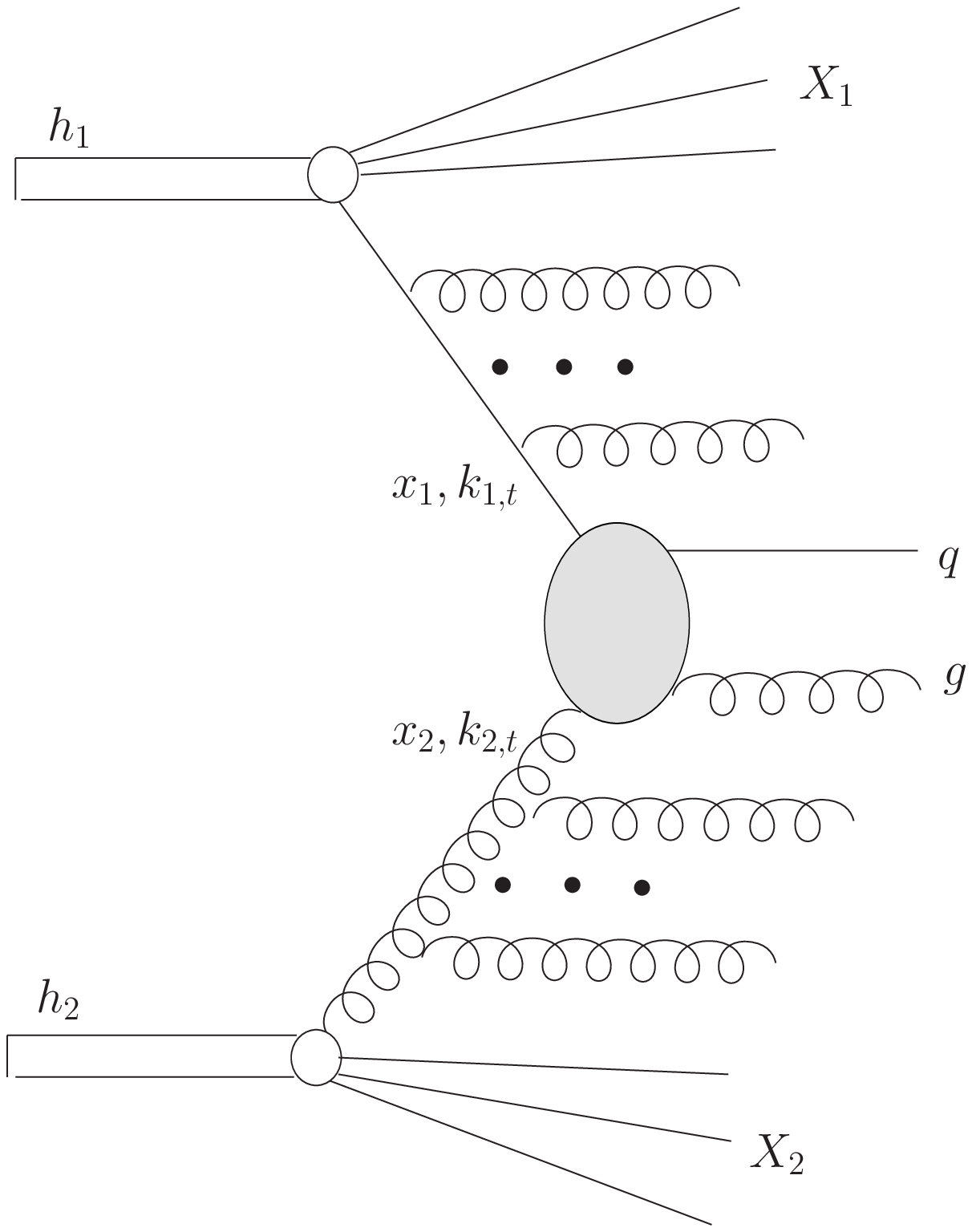} }
\caption{Diagrams for $k_t$-factorization approach to dijet
production.}
\label{fig:kt_factorization_dijets_diagrams}
\end{center}
\end{figure}



\begin{figure}    
\begin{center}
\resizebox{0.35\columnwidth}{!}{%
\includegraphics{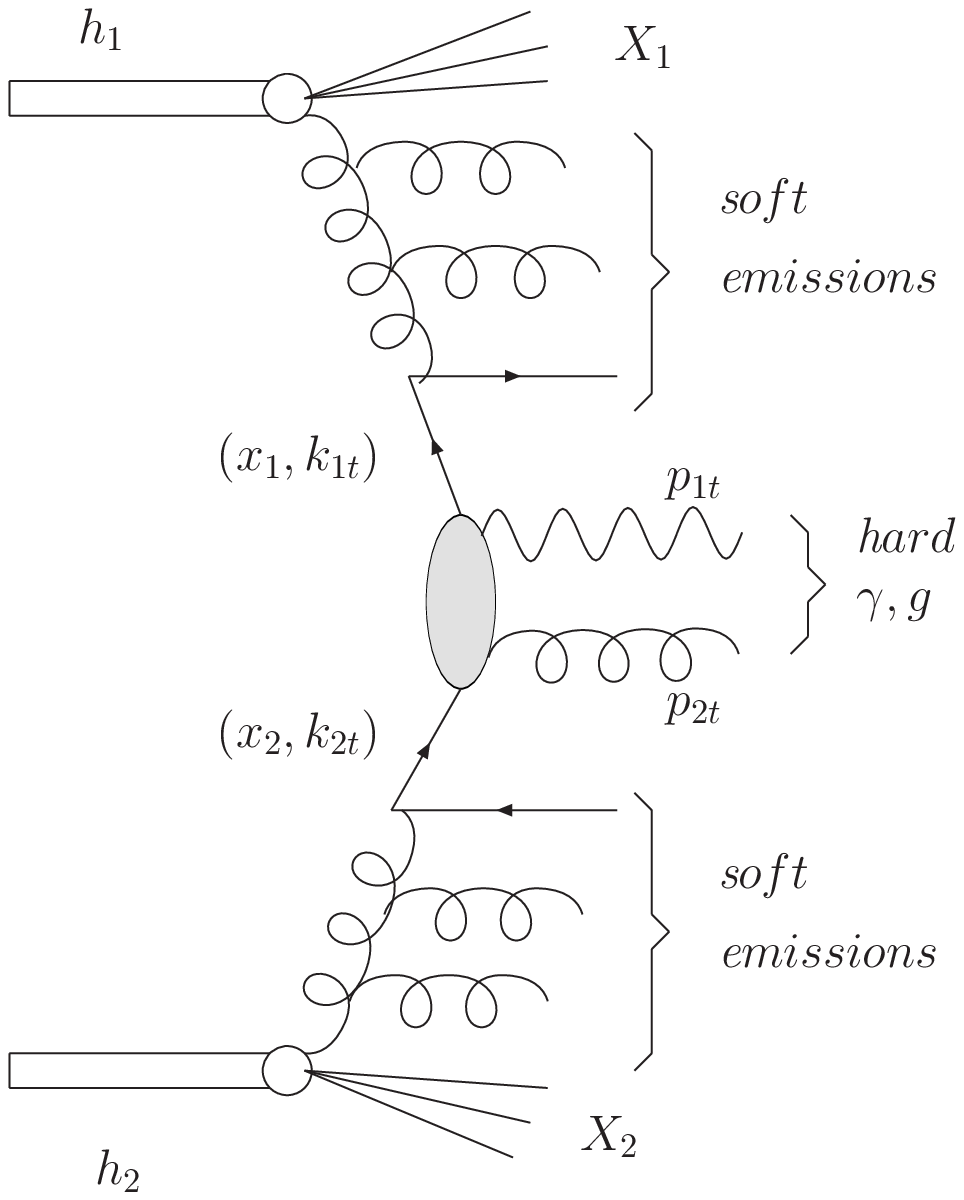} }
\resizebox{0.35\columnwidth}{!}{%
\includegraphics{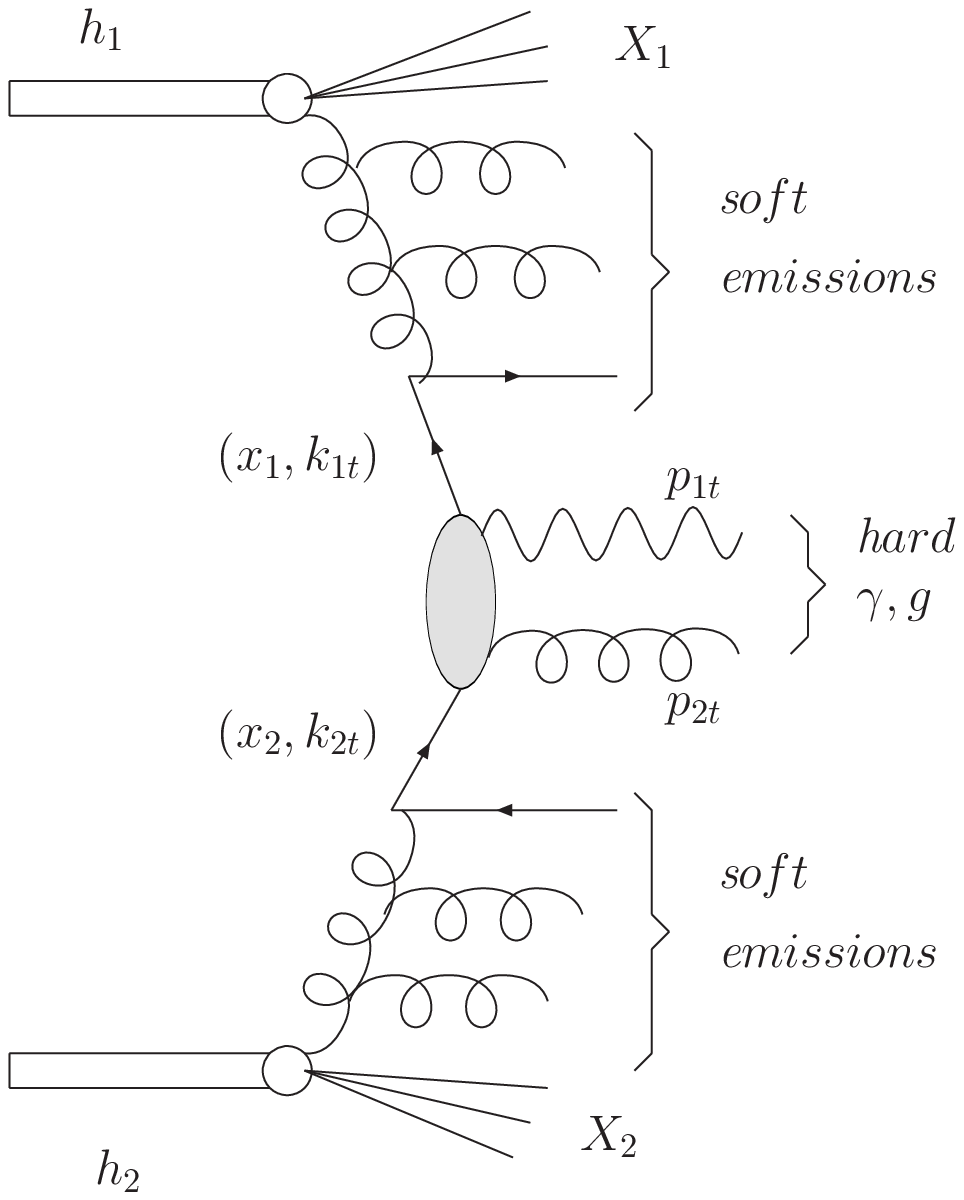} }
\resizebox{0.35\columnwidth}{!}{%
\includegraphics{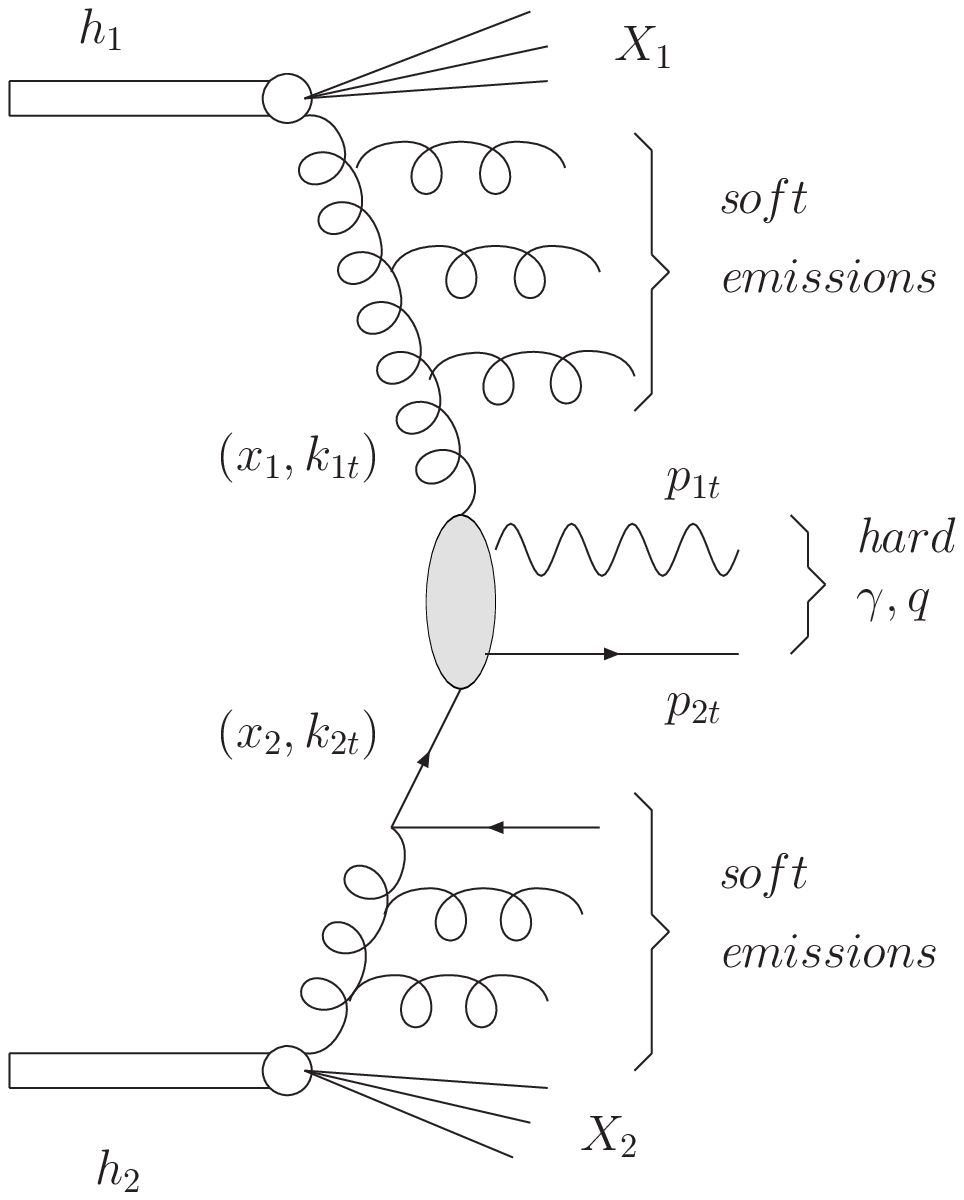} }
\resizebox{0.35\columnwidth}{!}{%
\includegraphics{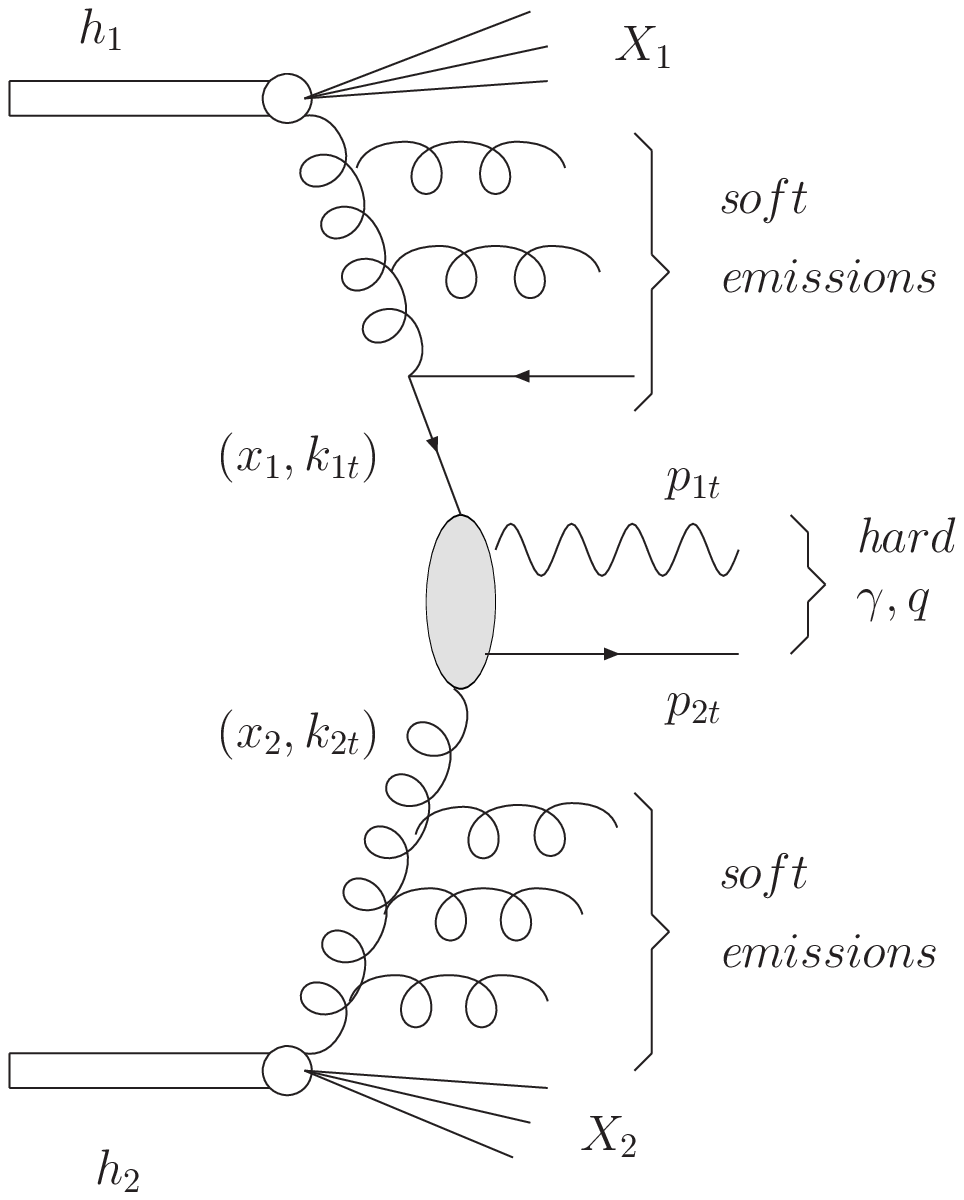} }
\caption{
Diagrams for $k_t$-factorization approach to photon-jet
correlations.}
\label{fig:kt_factorization_photonjet_diagrams}
\end{center}
\end{figure}



\begin{figure}    
\begin{center}
\includegraphics[width=6cm]{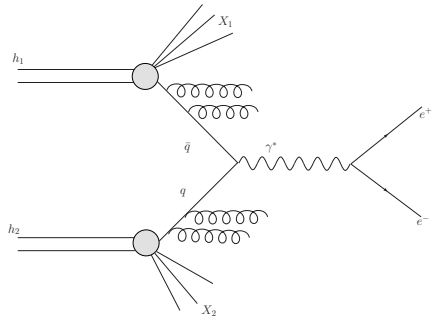}
\caption{Leading-order diagram for $k_t$-factorization approach to 
Drell-Yan dilepton production.}
\label{fig:kt_factorization_DY_diagrams}
\end{center}
\end{figure}


In the $k_t$-factorization approach the cross section for 
the production of a pair of elementary particles ($k$ and $l$) can be written as
\begin{eqnarray}
&&\frac{d\sigma(h_1 h_2 \rightarrow k l)}
{d^2p_{1,t}d^2p_{2,t}} 
= \sum_{i,j}
\int dy_1 dy_2
\frac{d^2 k_{1,t}}{\pi}\frac{d^2 k_{2,t}}{\pi}
\frac{1}{16\pi^2(x_1x_2s)^2} \nonumber \\
&&\overline{|{\cal M}(i j \rightarrow k l)|^2}
\delta^2(\overrightarrow{k}_{1,t}
+\overrightarrow{k}_{2,t}
-\overrightarrow{p}_{1,t}
-\overrightarrow{p}_{2,t})
{\cal F}_i(x_1,k_{1,t}^2){\cal F}_j(x_2,k_{2,t}^2) \; ,
\label{basic_formula}
\end{eqnarray} 
where
\begin{equation}
x_1 = \frac{m_{1,t}}{\sqrt{s}}\mathrm{e}^{+y_1} 
    + \frac{m_{2,t}}{\sqrt{s}}\mathrm{e}^{+y_2} \; ,
\end{equation}
\begin{equation}
x_2 = \frac{m_{1,t}}{\sqrt{s}}\mathrm{e}^{-y_1} 
    + \frac{m_{2,t}}{\sqrt{s}}\mathrm{e}^{-y_2} \; ,
\end{equation}
and $m_{1,t}$ and $m_{2,t}$ are so-called transverse masses
defined as $m_{i,t} = \sqrt{p_{i,t}^2+m^2}$, where $m$ is the mass
of the elementary particle.
In the following we shall assume that all partons are massless.
The objects denoted by ${\cal F}_i(x_1,k_{1,t}^2)$ and
${\cal F}_j(x_2,k_{2,t}^2)$ in the equation above are the unintegrated
parton distributions in hadron $h_1$ and $h_2$, respectively.
They are functions of longitudinal momentum fraction and transverse
momentum of the incoming (virtual) parton.
The sum is over all parton combinations leading to final
particles $k$ and $l$.

In Fig.\ref{fig:kt_factorization_ccbar_diagrams} I show the leading
diagrams for $c \bar c$ production. The heavy quark production
is considered as a flag process for $k_t$-factorization approach.
In Fig.\ref{fig:kt_factorization_dijets_diagrams} I show the diagrams
included for dijet correlations in Ref. \cite{SRS07}.
In Fig.\ref{fig:kt_factorization_photonjet_diagrams} I show similar
diagrams included for photon-jet correlations in Ref. \cite{PS07}.
Finally in Fig.\ref{fig:kt_factorization_DY_diagrams} I show basic 
leading-order diagram for Drell-Yan processes \cite{SS08}.
In the case of dilepton-jet correlations one needs to include
also higher-order processes (see \cite{SS08}).

The formula (\ref{basic_formula}) allows to study different types
of correlations. Here I shall limit to only few examples for
dijet and photon-jet correlations. More examples for $c \bar c$ 
were discussed by 
M. {\L}uszczak \cite{Luszczak_thisproceedings} and by 
G. \'Slipek for Drell-Yan processes \cite{Slipek_thisproceedings} at 
this workshop. The details concerning unintegrated gluon (parton) 
distributions can be found in original publications (see
e.g.\cite{LS06} and references therein).

\section{Selected results}
\label{results}


\begin{figure}    
\begin{center}
\resizebox{0.35\columnwidth}{!}{%
\includegraphics{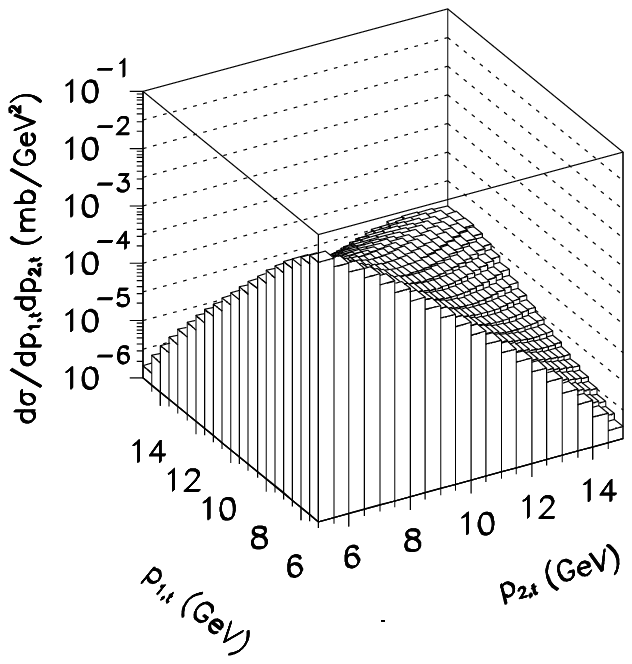}}
\resizebox{0.35\columnwidth}{!}{%
\includegraphics{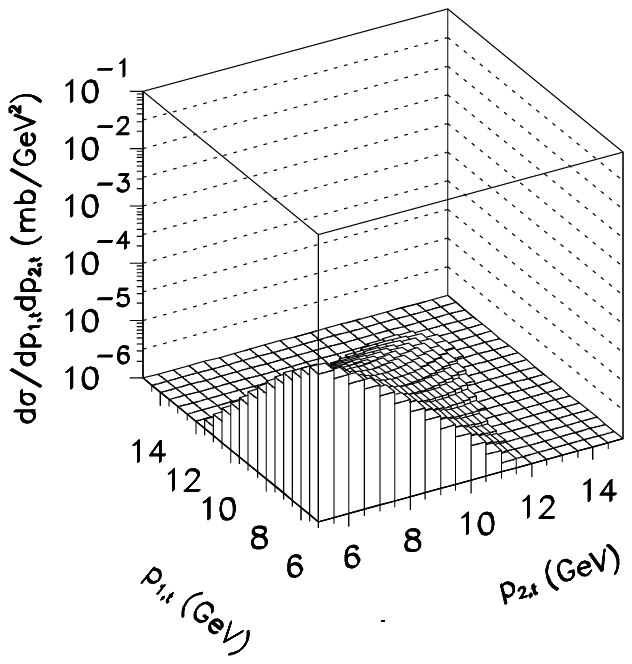}}
\resizebox{0.35\columnwidth}{!}{%
\includegraphics{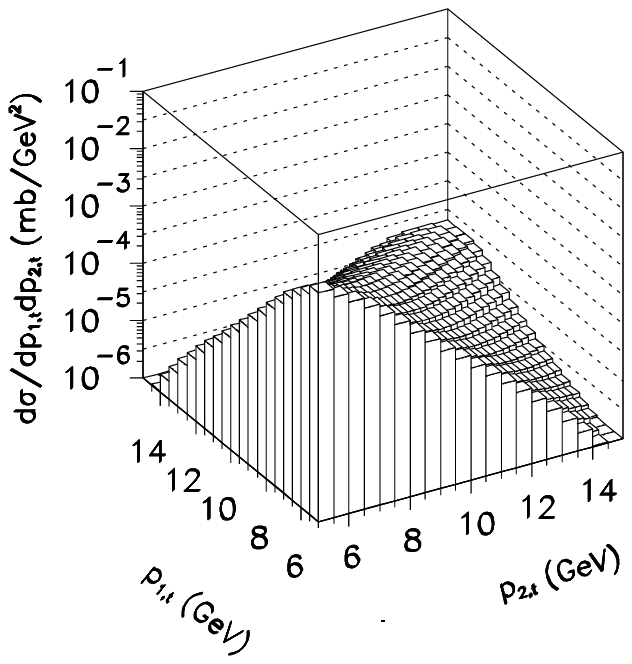}}
\resizebox{0.35\columnwidth}{!}{%
\includegraphics{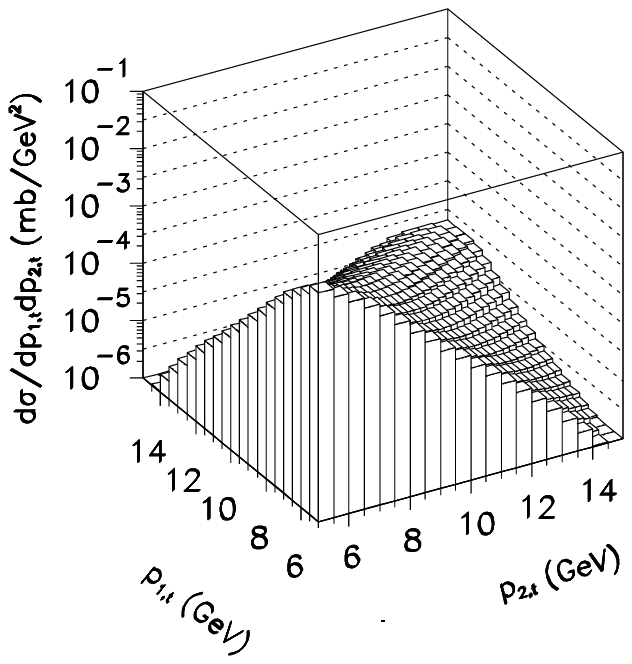}}
\caption{
Two-dimensional distributions in $p_{1,t}$ and $p_{2,t}$ 
for different subprocesses $gg \to gg$ (upper left)
$gg \to q \bar q$ (upper right), $gq \to gq$ (lower left)
and $qg \to qg$ (lower right). In this calculation $\sqrt{s}$ = 200 GeV 
and Kwieci\'nski UPDFs with exponential nonperturbative form factor
($b_0$ = 1 GeV$^{-1}$) and $\mu^2$ = 100 GeV$^2$ were used.
Here integration over full range of parton rapidities was made.}
\label{fig:p1tp2t_dijets_abcd}
\end{center}
\end{figure}


Let us start with dijet correlations.
As an example in Fig.\ref{fig:p1tp2t_dijets_abcd} we show 
two-dimensional maps of the cross section in 
$(p_{1,t},p_{2,t})$ for processes
shown in Fig.\ref{fig:kt_factorization_dijets_diagrams}.
Only very few approaches in the literature include both 
gluons and quarks and antiquarks.
In the calculation above we have used Kwieci\'nski UPDFs 
with exponential nonperturbative form factor 
\footnote{ $F(b) = \exp(-b^2/4 b_0^2)$ multiplies UPDFs 
in the impact parameter space and
is responsible for nonperturbative effects included 
in addition to perturbative effects embedded 
in the Kwieci\'nski evolution equations (for
more details see e.g. \cite{LS06}).}
($b_0$ = 1 GeV$^{-1}$),
and the factorization scale $\mu^2 = (p_{t,min}+p_{t,max})^2/4$ = 100 GeV$^2$.

\begin{figure} 
\begin{center}
\resizebox{0.50\columnwidth}{!}{%
\includegraphics{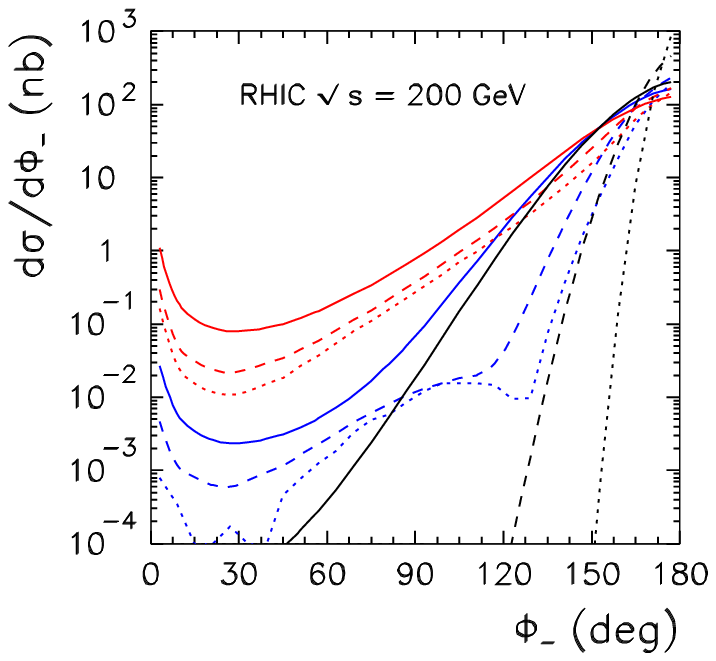}}
\caption{
(Color online) Azimuthal angle correlation functions at RHIC
energies for different scales and different values of $b_0$ 
of the Kwieci\'nski distributions.
The solid line is for $b_0$ = 0.5 GeV$^{-1}$, the dashed line is for
$b_0$ = 1 GeV$^{-1}$ and the dotted line is for $b_0$ = 2 GeV$^{-1}$.
Three different values of the scale parameters are shown: 
$\mu^2$ = 0.25, 10, 100 GeV$^2$ (the bigger the scale the bigger
the decorellation effect, different colors on line).
In this calculation  $p_{1,t}, p_{2,t} \in$ (5,20) GeV and
$y_1, y_2 \in$ (-5,5).
}
\label{fig:photon_jet_kwiecinski_scale}
\end{center}
\end{figure}

The second example is for photon-jet correlations.
In Fig.\ref{fig:photon_jet_kwiecinski_scale} we show the effect of
the scale evolution of the Kwieci\'nski UPDFs on the azimuthal angle
correlations between the photon and the associated jet.
We show results for different initial conditions 
($b_0$ = 0.5, 1.0, 2.0 GeV$^{-1}$). 
At the initial scale (fixed here as in the original
GRV \cite{PS07} to be $\mu^2$ = 0.25 GeV$^2$) there is 
a sizable difference of the results for different $b_0$. 
The difference becomes less and less pronounced when 
the scale increases.
At $\mu^2$ = 100 GeV$^2$ the differences practically disappear.
This is due to the fact that the QCD-evolution broadening of
the initial parton transverse momentum distribution is much 
bigger than the typical initial nonperturbative transverse 
momentum scale.


\begin{figure}
\begin{center}
    \includegraphics[width=10.0cm]{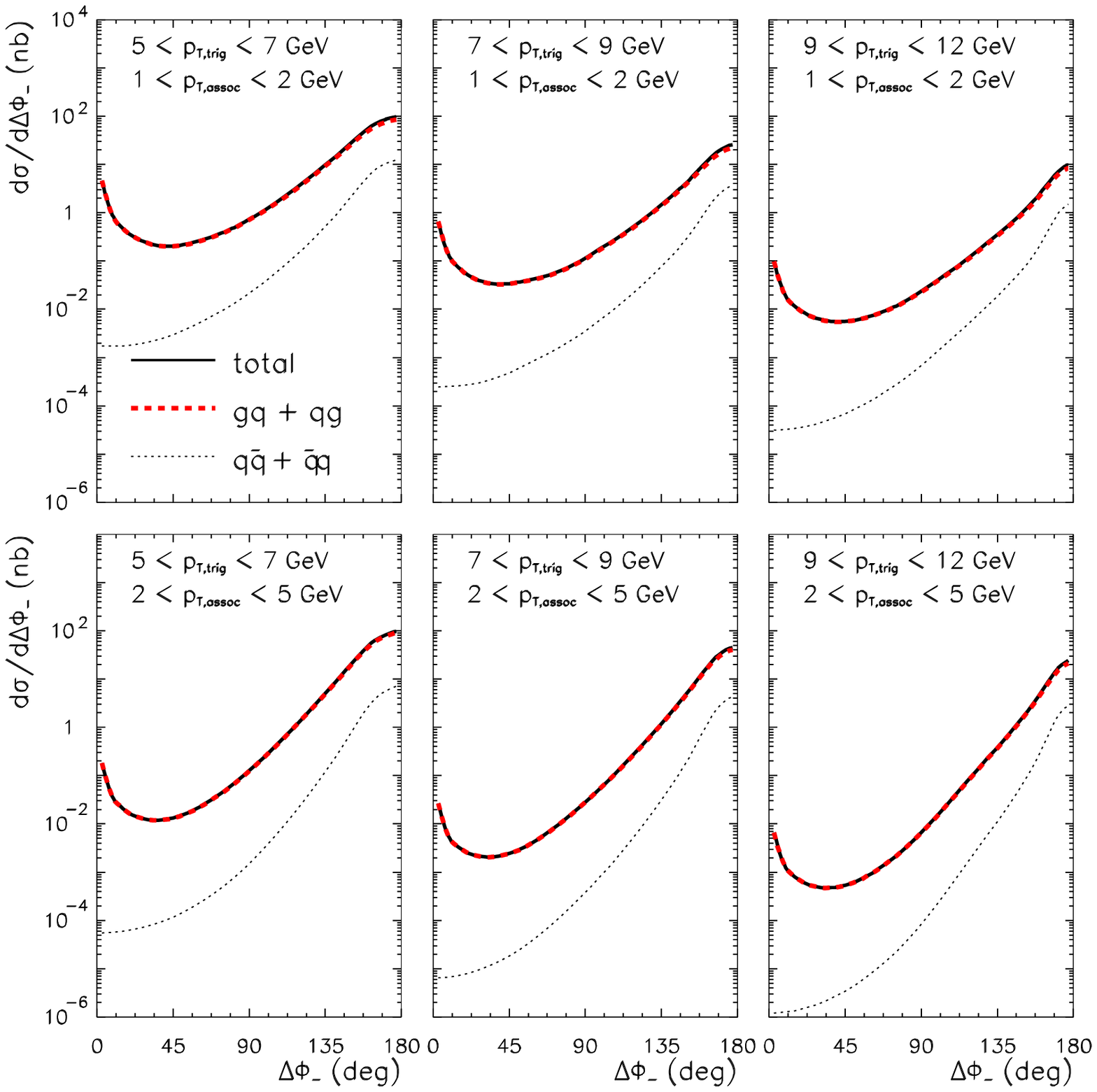}
\caption{Photon-hadron azimuthal correlations for different
windows of transverse momentum of the photon ($p_{T,trig}$)
and charged pion and kaon ($p_{T,assoc}$).
In this calculation $W$ = 200 GeV.
\label{fig:photon_hadron_correlations}}
\end{center}
\end{figure}


\begin{figure}
\begin{center}
    \includegraphics[width=6.0cm]{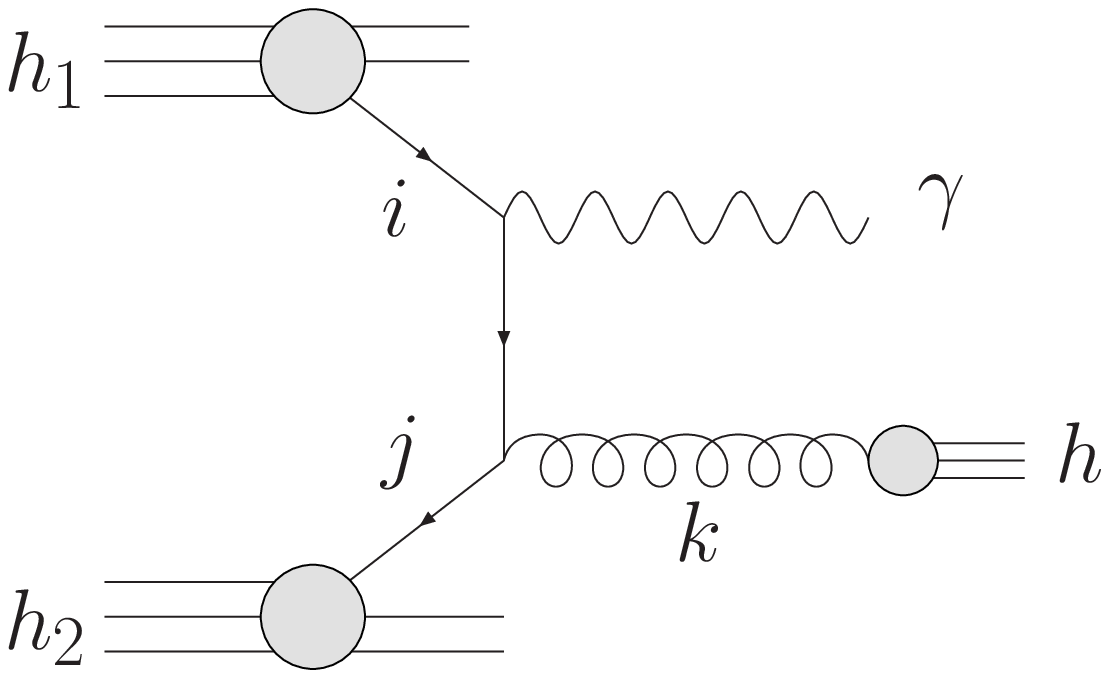}
\caption{
A schematic diagram illustrating associated production
of photon and hadron.
\label{fig:photon_hadron_diagram}
}
\end{center}
\end{figure}


Until now at the RHIC only photon-hadron correlation
were measured by the PHENIX collaboration \cite{Phenix}.
I show corresponding preliminary results of our 
calculation in Fig.\ref{fig:photon_hadron_correlations}.
The results are shown for different intervals
of $p_{T,tig}$ (photon) and $p_{T,assoc}$ (charged hadron).
The intervals correspond to the experimental cuts.
This calculation requires inclusion of fragmentation
functions (see Fig.\ref{fig:photon_hadron_diagram}). 
We have included both charged pions and charged
kaons in this calcultion. It was assumed that pions/kaons 
are emitted parallel to the jet (parton) direction.
The details will be shown elsewhere.
The distributions for different windows of transverse 
momenta are rather similar.

\section{Conclusions}
\label{conclusions}

Motivated by the recent experimental results of 
hadron-hadron correlations at RHIC I have discussed $c \bar c$, 
jet-jet, photon-jet and $e^+ e^-$ correlations.

In comparison to recent works on dijet production in the framework of
$k_t$-factorization approach, we have included two new mechanisms 
based on $gq \to gq$ and $qg \to qg$ hard subprocesses.
This was done using the Kwieci\'nski unintegrated parton
distributions.
We find that the new terms give significant contribution at RHIC energies.
In general, the results of the $k_t$-factorization approach depend
on UGDFs/UPDFs used, i.e. on approximation and assumptions made
in their derivation.

In the region of small transverse momenta of outgoing particles
the $k_t$-factorization approach is a good and efficient tool 
for the description of correlations \cite{SRS07}.
Rather different results are obtained with different UGDFs
which opens a possibility to verify them experimentally.
Consequences for particle-particle correlations, measured
recently at RHIC, require a separate dedicated analysis.

We have discussed also photon-jet correlation observables.
Up to now such correlations have not been studied experimentally.
As for the dijet case we have concentrated on the
region of small transverse momenta (semi-hard region) where
the $k_t$-factorization approach seems to be the most efficient and
theoretically justified tool.
We have calculated correlation observables for different unintegrated parton
distributions from the literature. Our previous analysis of
inclusive spectra of direct photons suggests that the Kwieci\'nski
distributions give the best description at low and intermediate
energies.
We have discussed the role of the evolution scale of the Kwieci\'nski
UPDFs on the azimuthal correlations. In general, the bigger the scale
the bigger decorrelation in azimuth is observed. When the scale
$\mu^2 \sim p_t^2$(photon) $\sim p_t^2$(associated jet)
(for the  kinematics chosen $\mu^2 \sim$ 100 GeV$^2$) is assumed, much bigger 
decorrelations can be observed than from the standard Gaussian smearing
prescription often used in phenomenological studies.

At RHIC one can measure jet-hadron correlations only for not too high
transverse momenta of the trigger photon and of the associated hadron.
This is precisely the semihard region discussed here.
In this case the theoretical calculations require inclusion of the
fragmentation process. This was done assuming independent parton
fragmentation method with the help of fragmentation functions
taken from the literature.

All the correlation observables have been studied for
RHIC or even lower energies. This can be repeated
in the future for LHC energies.


\end{document}